%% file: 00-CHIIR24.tex
\documentclass[sigconf,screen]{acmart}

\AtBeginDocument{%
  \providecommand\BibTeX{{%
    \normalfont B\kern-0.5em{\scshape i\kern-0.25em b}\kern-0.8em\TeX}}}

\input{00-macros}

\acmSubmissionID{13760.21}

\usepackage{soul}
\usepackage{wrapfig,lipsum,booktabs}
\usepackage{cancel}

\copyrightyear{2024} \acmYear{2024} \setcopyright{rightsretained} \acmConference[CHIIR '24]{Proceedings of the 2024 ACM SIGIR Conference on Human Information Interaction and Retrieval}{March 10--14, 2024}{Sheffield, United Kingdom}\acmBooktitle{Proceedings of the 2024 ACM SIGIR Conference on Human Information Interaction and Retrieval (CHIIR '24), March 10--14, 2024, Sheffield, United Kingdom}\acmDOI{10.1145/3627508.3638309} \acmISBN{979-8-4007-0434-5/24/03}
\begin{document}

\title[Walert]{Walert: Putting Conversational Search Knowledge into Action by Building and Evaluating a Large Language Model-Powered Chatbot}


\author{Sachin Pathiyan Cherumanal}
\orcid{0000-0001-9982-3944}
\affiliation{
\institution{RMIT University}
  \city{Melbourne}
  \country{Australia}
}
\email{s3874326@student.rmit.edu.au}

\author{Lin Tian}
\orcid{0000-0002-6889-2178}
\affiliation{
\institution{RMIT University}
  \city{Melbourne}
  \country{Australia}
}
\email{lin.tian2@student.rmit.edu.au}

\author{Futoon M.~Abushaqra}
\orcid{0000-0002-4892-5358}
\affiliation{
\institution{RMIT University}
  \city{Melbourne}
  \country{Australia}
}
\email{futoon.abu.shaqra@student.rmit.edu.au}

\author{Angel Felipe Magnoss\~ao de Paula}
\orcid{0000-0001-8575-5012}
\affiliation{
\institution{Universitat Polit\`ecnica de Val\`encia}
  \city{Valencia}
  \country{Spain}
}
\email{adepau@doctor.upv.es}

\author{Kaixin Ji}
\orcid{0000-0002-4679-4526}
\affiliation{
\institution{RMIT University}
  \city{Melbourne}
  \country{Australia}
}
\email{kaixin.ji@student.rmit.edu.au}

\author{Halil Ali}
\orcid{0009-0001-1335-9333}
\affiliation{
\institution{RMIT University}
  \city{Melbourne}
  \country{Australia}
}
\email{halil.ali@rmit.edu.au}

\author{Danula Hettiachchi}
\orcid{0000-0003-3875-5727}
\affiliation{
\institution{RMIT University}
  \city{Melbourne}
  \country{Australia}
}
\email{danula.hettiachchi@rmit.edu.au}

\author{Johanne R.~Trippas} 
\orcid{0000-0002-7801-0239}
\affiliation{
\institution{RMIT University}
  \city{Melbourne}
  \country{Australia}
}
\email{j.trippas@rmit.edu.au}

\author{Falk Scholer}
\orcid{0000-0001-9094-0810}
\affiliation{
\institution{RMIT University}
  \city{Melbourne}
  \country{Australia}
}
\email{falk.scholer@rmit.edu.au}

\author{Damiano Spina} 
\orcid{0000-0001-9913-433X}
\affiliation{
\institution{RMIT University}
  \city{Melbourne}
  \country{Australia}
}
\email{damiano.spina@rmit.edu.au}

\renewcommand{\shortauthors}{Pathiyan Cherumanal~et~al.}

\begin{abstract}
Creating and deploying customized applications is crucial for operational success and enriching user experiences in the rapidly evolving modern business world. 
A prominent facet of modern user experiences is the integration of chatbots or voice assistants. The rapid evolution of Large Language Models (LLMs) has provided a powerful tool to build conversational applications.
We present \walert, a customized LLM-based conversational agent able to answer frequently asked questions about computer science degrees and programs at RMIT University.
Our demo aims to showcase how conversational information-seeking researchers can effectively communicate the benefits of using best practices to stakeholders interested in developing and deploying LLM-based chatbots. These practices are well-known in our community but often overlooked by practitioners who may not have access to this knowledge. The methodology and resources used in this demo serve as a bridge to facilitate knowledge transfer from experts, address industry professionals' practical needs, and foster a collaborative environment. The data and code of the demo are available at~\url{https://github.com/rmit-ir/walert}. 
\end{abstract}





\begingroup
\mathchardef\UrlBreakPenalty=10000
\maketitle

\input{01-introduction}

\input{02-methodology}

\input{03-evaluation}
\input{04-conclusions}

\begin{acks}
\walert was designed and developed in the unceded lands of the  Wurundjeri and Boon Wurrung peoples of the eastern Kulin Nation. We pay our respects to their Ancestors and Elders, past, present, and emerging. This research is partially supported by the \grantsponsor{ARC}{Australian Research Council}{https://www.arc.gov.au/} (\grantnum{ARC}{DE200100064}, \grantnum{ARC}{CE200100005}) and is undertaken with the assistance of computing resources from RACE (RMIT AWS Cloud Supercomputing). We thank Amina Hossain and Santha Sumanasekara for their valuable contributions.
\end{acks}

\bibliographystyle{ACM-Reference-Format}


\bibliography{99-refs}


\end{document}

%% file: 00-macros.tex
\usepackage{pbalance}
\usepackage{xspace}
\usepackage{xcolor}
\usepackage{soul} 
\usepackage{siunitx}

\renewenvironment{quote}
  {\list{}{\rightmargin=0.4cm \leftmargin=0.4cm}%
   \item\relax}
  {\endlist}

\newcommand{\walert}{{Walert}\xspace}

\newcommand{\jt}[1]{\textcolor{blue}{\textbf{[[JT: #1]]}}} 
\newcommand{\ds}[1]{\textcolor{orange}{\textbf{[[DS: #1]]}}} 

\newcommand{\dha}[1]{\textcolor{cyan}{\textbf{[[DH: #1]]}}} 
\newcommand{\ap}[1]{\textcolor{red}{\textbf{[[AP: #1]]}}} 
\newcommand{\fs}[1]{\textcolor{purple}{\textbf{[[FS: #1]]}}} 
\newcommand{\kx}[1]{\textcolor{teal}{\textbf{[[KX: #1]]}}} 

\newcommand{\todo}[1]{\textcolor{red}{[[#1]]}}

\newcommand{\spc}[1]{\textcolor{red}{[{\bf SPC: #1}]}}

\renewcommand{\jt}[1]{}
\renewcommand{\ds}[1]{}
\renewcommand{\dha}[1]{}
\renewcommand{\ap}[1]{}
\renewcommand{\fs}[1]{}
\renewcommand{\kx}[1]{}
\renewcommand{\todo}[1]{}
\renewcommand{\spc}[1]{}

\usepackage{multirow}
\usepackage{enumitem}

%% file: 01-introduction.tex
\section{Introduction}
\label{sec:intro}

Conversational agents based on Large Language Models (LLMs) such as OpenAI's ChatGPT\footnote{\url{https://openai.com/}} provide many benefits to stakeholders, reducing the cost of various tasks by saving time and resources, especially for closed-domain questions such as translating responses into different formats, retrieving policy documents, and preparing legal document drafts~\cite{deroy2023ready}. However, two major concerns arise in this setting: 
\textit{(i)}~the risk of giving away sensitive data about the organization~\cite{ye2020chatbot, gobel2013information}; and 
\textit{(ii)}~the limited access to structured and comprehensible documentation might impede practitioners (e.g., data scientists without a strong background on information retrieval) grasp of the principles, theories, and best practices influencing product quality. Efforts have been taken in this direction of bringing together researchers and experts in conversational information-seeking with industry practitioners to deal with real-world problems -- as in the case of Amazon's Alexa Prize challenges \cite{agichtein2022alexa, yoelle2022taskbot,ram2018conversational}.

We aim to combine our diverse research expertise in machine learning, natural language processing, and information retrieval, to bring conversational search knowledge into action and address best practices for building an LLM-powered chatbot. Our goals are to explore the challenges and approaches for implementing a chatbot, compare various methods, and offer best practices in evaluation that stakeholders (not necessarily experts in conversational information seeking) can effectively use to assess and improve the quality of chatbot products.

Using a manually curated Frequently Asked Questions (FAQ) guide from RMIT University's School of Computing Technologies as a Knowledge Base (KB), we were able to implement a conversational agent, named \walert\footnote{The term ``Walert'' means ``possum'' in the native languages of the Woi Wurrung and Boon Wurrung peoples. Possum skin cloaks are essential to the Traditional Owners and Custodians of the land where the authors live and work. Our chatbot, Walert, is named as a tribute to this cultural heritage~\cite{wurrunggibiik2019,boonwurrung2016}.}, that allows potential future students to get answers to questions related to the computer science programs offered at RMIT University.\footnote{\url{https://www.rmit.edu.au/partner/hubs/race/news}} 

In this demo, the primary focus was to characterize the challenges associated with integrating LLMs into the development process of voice-based conversational information-seeking systems based on an existing KB. The process allowed us to address challenges related to 
\textit{(i)}~handling private/sensitive information by deploying our instance of an open-source LLM; 
\textit{(ii)}~monitoring the problem of hallucinations (i.e., the introduction of facts that are not true)~\cite{ji2023hallucinationSurvey} and generation of inaccurate information by using a human-in-the-loop approach~\cite{Gao2021HumanAIHybrid} and the inclusion of out of KB questions in our testbed; and 
\textit{(iii)}~evaluating the effectiveness of intent-based using Natural Language Understanding (NLU) and Retrieval-Augmented Generation (RAG), both at component and end-to-end levels. Our evaluation highlights shortcomings in recent RAG pipeline studies, particularly regarding the lack of ranking evaluation.

%% file: 02-methodology.tex
\section{Methodology}
\label{sec:methodology}

\begin{figure}[tp]
    \centering
    \includegraphics[width=0.8\linewidth]{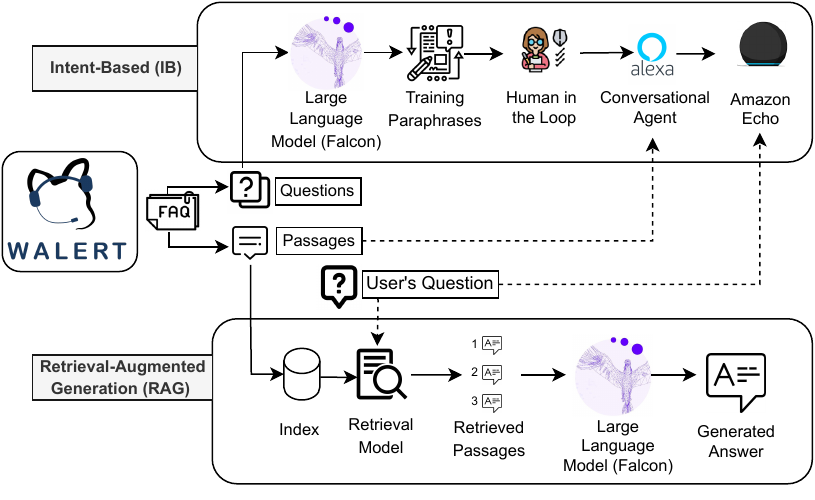}
    \caption{Overall architecture of the two approaches implemented in \walert{}: IB and RAG.}
    \label{fig:overview}
\end{figure}

Two different approaches were used to implement the prototype of our application: Intent-Based (IB) and RAG. The first one, IB, is suitable for structured and predictable interactions (i.e., pre-defined intents), while RAG chatbots retrieve information from a 
KB and are better for open domain and more dynamic conversations.
Figure~\ref{fig:overview} presents the overall framework for both approaches.

\raggedbottom

\subsection{Data Collection and Testbed}

We utilized a manually curated FAQ from RMIT University's School of Computing Technologies 
as a KB. The FAQ contains a wide range of common questions that incoming students ask regarding course offerings and academic programs related to computer science. 
Using the question-answer pairs in the FAQ, we generate a set of questions $Q$ and passages $P$.
The question set $Q = \{q_1, q_2, \ldots, q_n\}$  contains existing questions in the FAQ that have known answers (i.e., passages directly extracted from the FAQ), new questions with inferred answers (i.e., manually generated questions that have no direct answers but the answer can be inferred from multiple passages in the FAQ), and questions that do not have an answer in the KB (questions were manually generated by checking that they cannot be answered with the passages in the KB).

To simulate the scenario of different users expressing similar questions differently, we generated multiple semantically equivalent variations for each unique question initially included in the FAQ. The passages set $P = \{p_1, p_2, \ldots, p_m\}$ is a corpus of passages extracted from the FAQ, representing our KB. Finally, each question is associated with a gold answer from a set 
$A={\{a_1, \ldots, a_n\}}$. Each answer $a$ is obtained from one or more passages. 
We created a testbed that consists of relevance judgments at the passage level and gold answers for three types of questions:

\noindent \textbf{\textit{Questions with Known Answers (Known).}} These questions have a direct answer in the FAQ. Therefore, the corresponding passage $p$ to the answer $a$ in the FAQ is judged as \emph{Highly Relevant} ($\mbox{label} = 2$). All questions related to the same topic have the same passage judged as relevant, which is also the gold answer ($p=a$).

\noindent \textbf{\textit{Questions with Inferred Answers (Inferred).}} Questions that do not have a direct answer in the FAQ, but have an answer that can be extracted from the KB, i.e., from one or more passages. Passages that partially contain relevant information to answer the question are judged as \emph{Partially Relevant} ($\mbox{label} = 1$). The gold answer is manually generated by combining multiple passages.

\noindent \textbf{\textit{Out-of-Knowledge Base Questions (Out of KB).}} These questions cannot be answered with the information available in the KB -- even though, these questions are within the domain and likely to be asked. Therefore, there are no relevant passages, and the gold answer consists of communicating to the user that there is no information available to answer that question.

Table~\ref{tab:example questions} shows example passages and answers for each question type. We have 106 questions (including variations) and 120 passages. In our collection, 84 questions have known answers (passages directly extracted from the FAQ), 12 have inferred answers, and 10 do not have an answer in the KB. These three question types in our testbed provide a comprehensive evaluation that allows us to compare IB against RAG conversational approaches.

\input{02-tab-data}

In particular, using the questions with known answers, we can assess the system's ability to correctly respond to questions for which we know a passage in the KB contains the complete answer. By evaluating the effectiveness of the answers for the questions with inferred answers, we can assess the system's ability to generate answers by combining multiple (partially) relevant passages. Finally, including questions not covered in the FAQ helps assess the chatbot's ability to identify unanswerable questions -- which is a critical step in controlling hallucinations.

\subsection{Intent-Based (IB)}
\label{subsec:intent_walert}

The IB approach consists of a conversational model built using Amazon Alexa Skills~\cite{ram2018conversational} (upper part of Figure~\ref{fig:overview}). Each question in the FAQ is mapped to an \emph{intent} in the conversational model, i.e., one of the possible actions recognizable by the system.\footnote{\url{https://developer.amazon.com/en-US/docs/alexa/custom-skills/create-intents-utterances-and-slots.html}} Intuitively, this approach aims to optimize the correctness of the answers (high precision) but only handles a limited number of questions (low recall), i.e., those present in the FAQ (questions with known answers). Building effective intent recognition models requires multiple instances to train each intent. Since manually creating variations of training utterances is time-consuming, we experimented with using open-source LLMs to automatically create semantically equivalent variations of utterances (i.e., training data augmentation). In our case, these instances would be the semantic variations of the questions from the FAQ.
We deployed Falcon-7B~\cite{falcon40b} in Amazon SageMaker Studio with a \texttt{5xlarge-GPU} configuration\footnote{We explored different alternatives that would allow us to better understand the process of managing a privacy-aware LLM solution in-house to avoid the potential leaking of sensitive data by using third-party solutions. We found this alternative to be a good fit for our needs, also the most flexible for our future research, e.g., experiments involving fine-tuning.} to generate up to eight question variations for each intent (i.e., a question in the FAQ), using a zero-shot approach and the following prompt: \emph{``generate up to eight paraphrases of the following question: QUESTION''}. After manually inspecting the variations generated, we established a threshold and selected the top five. These variations were then used to train the conversational model. We also normalized the instances by resolving the acronyms (e.g., replacing CS with Computer Science) and used them along with the original questions for training, making it a total of six training instances per the intent of the conversational model. 
The answers associated with the original questions in the FAQ were used as responses returned for each question. The Alexa Skill was finally deployed in an Amazon Echo ($5^{th}$ generation) device, which allowed users to interact with the system in an audio-only setting. To enable this, we utilized the Automatic Speech Recognition and NLU features built into Amazon Alexa Skill.

 \subsection{Retrieval-Augmented Generation (RAG)}

In contrast to the precision-oriented IB approach, we sought to investigate a more open-ended methodology known as 
RAG~\cite{lewis2020retrieval}. Here, we cover two use cases: \textit{(i)}~instances where the questions the user raises vary from what is available in the FAQ, and \textit{(ii)}~scenarios in which questions can only be answered using abstractive multi-document summarization from multiple passages in the KB. The RAG approach consists of two main stages (bottom part of Figure~\ref{fig:overview}). The initial stage involves retrieving potential passages that may contain the answer, while the second stage involves generating a summary from the top-$K$ retrieved passages.

For the retrieval model, we experimented with two approaches: \textit{(i)}~Okapi BM25~\cite{robertson1995okapi} with default parameters ($k1 = 1.2$; $b = 0.75$) and \textit{(ii)}~dense retrieval using Dense Passage Retrieval (DPR)~\cite{karpukhin2020dpr} implementations in the \texttt{pyserini} toolkit~\cite{lin201pyserini}. To generate a summary from the top-$K$ retrieved passages, we used the same LLM that was used to generate semantically equivalent variations of the questions for the IB approach, i.e., \texttt{falcon-7b-instruct} (refer Section~\ref{subsec:intent_walert}) along with the following prompt to generate the summaries:

    \begin{quote}
    \it
    Generate an answer to be synthesized with text-to-speech for a virtual assistant, the answer should be based on the retrieved documents for the following question. If the retrieved documents are not related to the question, then answer NA.\newline
    [QUESTION + LIST OF $k$ PASSAGES]
    \end{quote}

We experimented with three top-heavy $k$ cutoffs: 1 (which is comparable to IB), 3, and 5 top retrieved passages.

%% file: 02-tab-data.tex
\begin{table*}[tp]
\caption{Examples of questions, relevant passages and gold answers in our testbed.} 
\label{tab:example questions}  
\small
\begin{tabular}{p{2cm}p{3.5cm}p{7cm}p{3cm}}
\toprule
 \textbf{Question Type}          & \textbf{Question}     & \textbf{Passage(s)}   & \textbf{Answer}    \\
\midrule
Questions with Known Answers    & 
Is the transfer from Associate Degree to Bachelors automatic? 
& No, it is not. You are required to apply when you are closer to the completion of the Associate Degree. ({\it Highly Relevant}) & No, it is not. You are required to apply when you are closer to the completion of the Associate Degree. \\

\midrule
Questions with Inferred Answers &
What does the final year of Computer Science (CS) program include? 
& (1) 
[\ldots]
Software Engineering (SE) students will do another large in-house project and more SE electives, while CS students will do a slightly smaller project and a few more core [\ldots].  ({\it Partially Relevant})
(2) [\ldots]
students are required by RMIT rules to do a capstone project in their final year. [\ldots] with an industry partner
[\ldots] ({\it Partially Relevant})
& It includes a small capstone project with a supervisor that work with an industry partner, as well as a few more core courses and electives. \\ 
\midrule
Out-of-KB Questions & 
When does the application for program transfer open?  
& Not available  ({\it No Relevant Passages})   &             I'm sorry, I don't have an answer.  \\ 
\bottomrule 
\end{tabular}
\end{table*}

%% file: 03-evaluation.tex
\section{Quantitative Evaluation}
\label{sec:Testing}

The test collection created from the knowledge base described in Section~\ref{sec:methodology}, allows us to apply effectiveness evaluation practices to compare our proposed approaches at both the component and end-to-end levels. Table~\ref{tab:results} displays results for IB and RAG approaches using evaluation measures across two dimensions: \textit{(i)}~retrieval effectiveness (Normalized Discounted Cumulative Gain, NDCG~\cite{jarvelin2002ndcg}) and \textit{(ii)}~natural language generation (BERTScore~\cite{zhang2019bertscore} and ROUGE-1~\cite{lin2004rouge}). ROUGE and BERTScore have been used to quantify hallucinations in LLMs automatically~\cite{ji2023hallucinationSurvey}. All the results in Table~\ref{tab:results} have been tested for statistical significance using Tukey's HSD and significance level $\alpha = 0.01$. Below, we discuss the results across two dimensions separately.

\input{03-tab-results}

\noindent \textbf{Retrieval Effectiveness.} When it comes to retrieving a response, the IB approach only retrieves one passage (i.e., the response associated with the recognized intent), whereas RAG approaches retrieve multiple passages for a given question (and the final response is generated using the top-$k$ passages). In our evaluation, we explore top-heavy cutoffs $k$=\{1,3,5\} to minimize the risk of hallucinations.
Table~\ref{tab:results} shows that, for the Known questions, IB and RAG using DPR have comparable performance in terms of NDCG@1 and RAG approaches perform better with more aggressive ranking truncation ($k=1$). For the Inferred questions, RAG approaches outperform IB (which can only return, at most, a passage partially relevant to the question). RAG approaches benefit from more context, and BM25 with $k=3$ obtains the highest NDCG score. In terms of out-of-KB questions, IB performs substantially better than RAG approaches, being able to identify 80\% of the unanswerable questions. RAG-based approaches fail by attempting to generate an answer for most of the questions, which means that it is likely to hallucinate instead of not warning the user about the lack of information in the KB.

\noindent \textbf{End-to-end evaluation.} BERTScore and ROUGE-1 scores indicate that IB performs significantly better for the Known questions, whereas RAG approaches are likely to generate better answers for the Inferred questions. It is worth noting that the LLM tend to benefit from having less context (i.e., fewer passages in the prompt), achieving higher BERTScore and ROUGE-1 scores for lower cutoffs for both RAG approaches.  Results also corroborate that IB performs significantly better than RAG approaches for Out of KB questions.

%% file: 03-tab-results.tex
\newcommand{\customsize}{\fontsize{6.5}{9.5}\selectfont}

\begin{table*}[th]
\caption{Quantitative evaluation of the retrieval phase (NDCG) and generated answers (BERTScore and ROUGE-1), broken down by type of questions. Effectiveness for Out of KB base questions is reported with the percentage of empty rankings / ``I don't have an answer'' responses. Boldface indicates the best score for each measure and $^*$ indicates statistically significant differences against all the other approaches according to Tukey's HSD and significance level $\alpha=0.01$. \label{tab:results}}
\sisetup{
table-auto-round=true,
detect-weight=true,
table-format=1.3
}
\small
    \begin{tabular}{p{2.3cm}S[table-format=1]SSSSSSS[table-format=2.2]SS}
    \toprule
        \multirow{2}{*}{\it Approach} &  {\it Retrieval} &  \multicolumn{3}{c}{\it Known (84 Questions) } & \multicolumn{3}{c}{\it Inferred (12 Questions)} & \multicolumn{3}{c}{\it Out of KB (10 Questions)} \\
        \cmidrule(l){3-5}
        \cmidrule(l){6-8}
        \cmidrule(l){9-11}
        &                                      {\it Cutoff $k$ }                       &{\it NDCG} & {\it BERTScore} & {\it ROUGE-1} & {\it NDCG}&  {\it BERTScore} & {\it ROUGE-1} &{\it \% Unanswered } & {\it BERTScore} & {\it ROUGE-1} \\
        \midrule
        Intent-Based (IB) &  &   0.6429                                  & \bfseries 0.771{$^*$} & \bfseries 0.671{$^*$} & 0.0833 & 0.332{$^*$} & 0.0616 & \bfseries 80 & \bfseries 0.866{$^*$} & \bfseries 0.813{$^*$} \\
        \midrule
        \multirow{3}{*}{RAG (BM25 + Falcon)} & 1 & 0.5119 &
                                                                                            0.536 & 0.179 & 0.1667 & 0.493 & 0.185 & 0& 0.336 & 0.0447\\
        & 3 & 0.4912 &
                                                                                            0.543 & 0.209 & 0.2556 &0.447 & 0.106 & 10 &0.293 & 0.0537\\ 
        & 5 & 0.4733 &
                                                                                            0.543 & 0.209 & \bfseries 0.3291 &0.447 & 0.106 & 10 &0.293 & 0.0537\\
        \midrule
        \multirow{3}{*}{RAG (DPR + Falcon)} & 1 & \bfseries 0.6905 &                                        0.545 & 0.193 & 0.25 & \bfseries 0.513 & \bfseries 0.192 & 10 &0.34 & 0.0475\\
                                             & 3 & 0.6119 &                                        0.564 & 0.244 & 0.235 &0.476 & 0.123 & 20 &0.311 & 0.0453\\
                                             & 5 & 0.5812 &                                        0.564 & 0.244 & 0.235  &0.476 & 0.123 & 20 &0.311 & 0.0453\\
        \bottomrule
    \end{tabular}
\end{table*}

%% file: 04-conclusions.tex
\section{Impact and Future Work} 
\label{sec:conclusion}

We built \walert, a conversational agent that answers FAQs about programs of study offered in the School of Computing Technologies at RMIT University. 
The IB approach, deployed on an Amazon Echo device, was showcased as a demo at the university's Open Day in August 2023 where potential future students learned about the use of LLM-based conversational systems and its risks and limitations. The demo, which was also showcased to visiting high-school students in September 2023, generated university-wide interest and connections, including the IT service team building a university-wide solution.

There are several limitations that we are aiming to address in future work. The current demo relies upon a limited knowledge base (i.e., a manually curated FAQ). We aim to reproduce our methodology with a more extensive set of documents, including brochures and internal web pages related to the delivery of CS programs. Further research on evaluation measures (beyond BERTScore and ROUGE) is needed to evaluate the validity of generated responses. We aim to explore other evaluation measures, including those for truncated rankings~\cite{amigo2022ranking} and other dimensions of LLM-based conversational systems~\cite{sakai2023swan}. Finally, we plan to deploy RAG approaches to perform online experimentation.

The process of building \walert helped us not only to share complementary knowledge across our group but also to facilitate knowledge translation within the university. We believe the approach and the lessons learned can help other researchers aiming to bridge the gap between experts and practitioners interested in building (and testing) LLM-based conversational information-seeking systems.